\documentclass[]{aastex631}
\newcommand{\roma}[1]{\uppercase\expandafter{\romannumeral#1}}
\newcommand{\speed}[1]{#1\,km\,s${}^{-1}$}

\newcommand{\acc}[1]{#1 km~s${}^{-2}$}
\newcommand{\nfig}[1]{Figure~\ref{#1}}

\newcommand{\uu}{$^\prime$}
\usepackage{amsmath}
\usepackage{subfigure}
\usepackage{hyperref}%

 \usepackage{url}
\usepackage{graphicx}
\usepackage{natbib}
\usepackage{amssymb,txfonts}
\usepackage{multirow}
\usepackage{array}
\usepackage{sidecap}
\usepackage{hyperref}
\usepackage{epstopdf}
\usepackage{footnote}
\usepackage{tabularx}
\usepackage{booktabs}

\usepackage{CJK}

\hypersetup{
 bookmarks=true, 
 unicode=false, 
 pdftoolbar=true, 
 pdfmenubar=true, 
 pdffitwindow=false, 
 pdfstartview={FitH}, 
 pdftitle={My title}, 
 pdfauthor={Author}, 
 pdfsubject={Subject}, 
 pdfcreator={Creator}, 
 pdfproducer={Producer}, 
 pdfkeywords={keyword1} {key2} {key3}, 
 pdfnewwindow=true, 
 colorlinks=true, 
 linkcolor=blue, 
 citecolor=blue, 
 filecolor=blue
 urlcolor=blue
 }

\shorttitle{On the origin of a broad QFP wave train: unwinding jet as the driver}
\shortauthors{Zhou et al.}

\newcommand{\xp}[1]{{{\color{black}{#1}}}}
\newcommand{\zxp}[1]{{{\color{black}{#1}}}}
\newcommand{\xpz}[1]{{{\color{black}{#1}}}}
\newcommand{\zx}[1]{{{\color{black}{#1}}}}

\begin{document}
\begin{CJK*}{UTF8}{gbsn}
\title{On the origin of a broad QFP wave train: unwinding jet as the driver}

\correspondingauthor{Xinping Zhou}
\email{xpzhou@sicnu.edu.cn}

\correspondingauthor{Yuandeng Shen }
\email{ydshen@hit.edu.cn}

\author[0000-0001-9374-4380]{Xinping Zhou }
\affiliation{ College of Physics and Electronic Engineering, Sichuan Normal University, Chengdu 610068, People's Republic of China}
\affiliation{State Key Laboratory of Space Weather, Chinese Academy of Sciences, Beijing 100190}

\author[0000-0003-0880-9616]{Zehao Tang}
\affiliation{Yunnan Observatories, Chinese Academy of Sciences, Kunming 650216, People's Republic of China}

\author[0000-0001-8318-8747]{Zhining Qu}
\affiliation{ College of Physics and Electronic Engineering, Sichuan Normal University, Chengdu 610068, People's Republic of China}

\author[0000-0002-1134-4023]{Ke Yu}
\affiliation{ College of Physics and Electronic Engineering, Sichuan Normal University, Chengdu 610068, People's Republic of China}

\author[0009-0005-5300-769X]{Chengrui Zhou}
\affiliation{Yunnan Observatories, Chinese Academy of Sciences, Kunming 650216, People's Republic of China}

\author{Yuqi Xiang}
\affiliation{ College of Physics and Electronic Engineering, Sichuan Normal University, Chengdu 610068, People's Republic of China}

\author[0000-0001-9540-5235]{ Ahmed Ahmed Ibrahim}
\affiliation{Department of Physics and Astronomy, College of Science, King Saud University, P.O. Box 2455, 11451 Riyadh, Saudi Arabia}

\author[0000-0001-9493-4418]{Yuandeng Shen }
\affiliation{Shenzhen Key Laboratory of Numerical Prediction for Space Storm, School of Aerospace, Harbin Institute of Technology, Shenzhen 518055, China}

\begin{abstract}
Large-scale extreme-ultraviolet (EUV) waves commonly exhibit as single wavefront and are believed to be caused by coronal mass ejections (CMEs). Utilizing high spatiotemporal resolution imaging observations from the Solar Dynamics Observatory, we present two sequentially generated wave trains originating from the same active region: a narrow quasiperiodic fast-propagating (QFP) wave train that propagates along the coronal loop system above the jet and a broad QFP wave train that travels along the solar surface beneath the jet. The measurements indicate that the narrow QFP wave train and the accompanying flare's quasiperiodic pulsations (QPPs) have nearly identical onsets and periods. This result suggests that the accompanying flare process excites the observed narrow QFP wave train. However, the broad QFP wave train starts approximately 2 minutes before the QPPs of the flare, but consistent with the interaction between the unwinding jet and the solar surface. Moreover, we find that the \zx{period of the broad QFP wave train, approximately 130\,s, closely matches that of the unwinding jet}. This period is significantly longer than the 30\,s period of the accompanying flare's QPPs. Based on these findings, we propose that the intermittent energy release of the accompanying flare excited the narrow QFP wave train confined propagating in the coronal loop system. The unwinding jet, rather than the intermittent energy release in the accompanying flare, triggered the broad QFP wave train propagating along the solar surface. 

\end{abstract}


\section{Introduction} \label{sec:intro}

Hydromagnetics waves, as a fundamental physical phenomenon, can be utilized to infer the physical parameters of the medium they propagate. Consequently, \zxp{they offer a mean} to diagnose coronal physical parameters, such as magnetic field strength, utilizing seismological diagnostic techniques \citep{1970PASJ...22..341U,1984ApJ...279..857R}, and provide a possibility to explore the plasma density, temperature, as well as magnetic field strength inside the coronal hole using the lensing effect of the waves \citep{2024NatCo..15.3281Z,2022ApJ...930L...5Z}.

In the last half-century, large-scale extreme ultraviolet (EUV) waves have attracted many researchers to investigate them \citep{1998GeoRL..25.2465T}. Their physical nature and driven mechanism have finally been unified after many years of debate, namely, they are a fast-mode magnetosonic wave driven by the lateral expansion of the accompanying coronal mass ejections (CMEs) \citep{2001JGR...10625089W,2002ApJ...572L..99C,2011ApJ...738..160M,2012ApJ...746...13L,2018ApJ...864L..24L,2019ApJ...870...15L,2020MNRAS.493.4816M,2022ApJ...928...98H,ChenReview}. It is important to note that small-scale EUV waves, which are not associated with CMEs, have also been observed. However, these waves's sizes are relatively small, and the number of observations is limited. For instance, the small-scale EUV wave triggered by breakout reconnection in the fan-spine topology has been documented \citep{2016ApJ...828...28K}. \cite{2017ApJ...851..101S,2018ApJ...860L...8S,2018MNRAS.480L..63S} and \cite{2020ApJ...894..139Z} found that the waves can be driven by the fast expansion of the lower coronal loops associated with the failed filament eruption or by the impingement of the associated coronal jets. Additionally, \cite{2023A&A...674A.167L} proposed that the expansion of the newly formed loops can also be one of the driving sources of the small-scale EUV wave, resembling the slingshot mechanism in the eruption of the coronal jets \citep{2015ApJ...804...88S,2021NatAs...5...54A}. At the same time, some researchers find that the small-scale EUV waves can be directly driven by the ejection of a blowout surge or jet \citep{2012A&A...541A..49Z,2013ApJ...764...70Z,2021ApJ...911...33C,2023ApJ...953..171H}. As pointed out by \cite{2018ApJ...860L...8S}, the small-scale EUV wave driven by the sudden loop expansion, the surges or jets have a shorter lifetime than those with typical hour-long lifetime excited by the CMEs \citep[see the reviews][and references therein]{2014SoPh..289.3233L,2015LRSP...12....3W}. For convenience, the large- and small-scale EUV waves are collectively referred to as classical EUV waves.

In the past, the classical EUV waves were consistently perceived as a single-pulse phenomenon, resulting in little attention being paid to their possible periodicities. Recently, two new types of EUV wave trains with multiple wavefronts, captured by Atmosphere Imaging Assembly \citep[AIA;][]{2012SoPh..275...17L} onboard the Solar Dynamics Observatory \citep[SDO;][]{2012SoPh..275....3P}, have rekindled interest in the study of the EUV wave, these are, narrow \citep{2011ApJ...736L..13L,2012ApJ...753...53S,2013A&A...554A.144Y,2022ApJ...941...59Z} and broad \citep{2012ApJ...753...52L,2017ApJ...844..149K,2019ApJ...873...22S,2021SoPh..296..169Z,2022A&A...659A.164Z,2022ApJ...930L...5Z,2024SCPMA..6759611Z}  quasiperiodic fast-propagating (QFP) wave trains, albeit only in sporadic cases. \xp{For detailed information regarding their respective features, please refer to Table 1 of the review article by \cite{2022SoPh..297...20S}}. The significant difference between them and the classical EUV waves is that these two types of waves have multiple wavefronts. The narrow QFP wave train, first reported by \cite{2011ApJ...736L..13L}, propagates along the coronal loops system with a speed ranging from \speed{300} to \speed{2400} in a confined manner \citep{2014SoPh..289.3233L,2022SoPh..297...20S}. Generally, they are believed to be triggered by the intermittent energy release in the accompanying flare \citep{2011ApJ...736L..13L,2012ApJ...753...53S,2022ApJ...941...59Z}. The broad QFP wave train, similar to the classical EUV wave propagating along the solar surface, will display the refraction, transmission as well as total reflection when they interact with various coronal structures, such as filaments, coronal loops, and coronal holes \citep{2012ApJ...753...52L,2019ApJ...873...22S,2022A&A...659A.164Z,2022ApJ...930L...5Z,2024arXiv240418391Z}. Notably, utilizing the theoretical results by \cite{2021A&A...651A..67P}, \cite{2022A&A...659A.164Z} first certified a broad QFP wave train \emph{total reflected} from the boundary of a south coronal hole, marking the last piece of the puzzle of the properties of the true wave to be found. According to the statistic by \cite{2022SoPh..297...20S}, the speed of the broad wave train is in the range of 370-\speed{1100}, which is consistent with that of the classical EUV with a speed of 200-\speed{1500}. Although many characteristics of the broad QFP wave train, as mentioned above, are similar to the classical EUV wave, the driven mechanism for the latter is hardly used to explain the formation of the broad QFP wave train. The main reason is that it is difficult for a single energy-releasing process, such as the eruption CME, to stimulate the wave train with multiple wavefronts. There are different insights into the trigger mechanisms of the few observed broad QFP waves. For example, \cite{2022ApJ...939L..18S} believe that the broad wave train might be triggered by the successive stretching of the magnetic field lines during the solar eruption, where the speed ratio of the inner edge of the wave train and the wave train is approximately one-third. Their results align with the simulation's expectations by \cite{2002ApJ...572L..99C}. \cite{2021ApJ...911L...8W} reproduced the broad QFP wave train with a dome shape propagating perpendicular to the magnetic field lines with a speed of about 550-\speed{700} associated with \zxp{a failed filament eruption}. Their result indicates that CME is not necessarily involved in broad QFP wave train excitation. This finding has been certified from the observation view. For instance, \cite{2022A&A...659A.164Z} and \cite{2022ApJ...930L...5Z} found that the broad QFP wave train share a common period with the associated flare. Thus, they propose that the nonlinear energy release processes in the accompanying flare should drive this QFP wave. Although the discovery of these two types of waves enriches the observation feature of the coronal wave, it also implies that the triggering mechanism of the EUV wave is still an open question. \xp{Notability, sometimes classical EUV waves can be captured when they encounter the coronal loops system, resulting in these waves propagate inside it \citep{2021SoPh..296..169Z,2024ScChE..67.1592L}. In this process, the wave speed will be significantly increased due to the wave mode conversion \citep{2016GMS...216..381C,2018ApJ...863..101C,2024NatCo..15.2667K}. Unfortunately, there is no observational evidence yet of broad QFP wave train being captured by coronal loops. }

In this letter, we report, for the first time, the direct observational evidence of the broad QFP wave train driven by the unwinding jet utilizing the SDO/AIA high \xp{spatiotemporal} resolution imaging data. In contrast, the intermittent energy release of the accompanying flare triggered the narrow QFP wave train. The layout of the remaining paper is as follows: Section\,\ref{se:results} gives the observational analysis and results; Section\,\ref{se:discussion} presents the discussion and conclusion.

\section{Results}
\label{se:results}
\subsection{The unwinding jet}
The event occurred close to the western disk limb of the NOAA AR 11149 on 2011 January 27, associated with a B6.6 class flare, whose start, peak, and end time \xp{were} 08:40\,UT, 08:50\,UT, and 08:53\,UT, respectively. An overview of the eruption source region dominated with some open loops, highlighted with the green extrapolated magnetic field lines, using the Potential Field Source \citep[PFSS;][]{2003SoPh..212..165S} mode, is presented in \nfig{fig:overview} (a). Under the loops system, a mini filament lay prone under it. The snapshots in \nfig{fig:overview} (b)-(i) show the evolution of the jet using AIA 304 \AA\ and 131 \AA\ \xp{raw images}. This jet displays a straightforward evolution process in the plane of the sky, which is advantageous for conducting a detailed study of its rotation process. From the animation of \nfig{fig:overview}, one can find that the jet started at about 08:35\,UT and ended at about 08:55\,UT, exhibiting apparent rotation motions around the jet's center axis, during its rising period. Beginning from 08:38\,UT, \zxp{the jet is started to unwind, and \xpz{its body} became brighter}, marked with the white arrows in \nfig{fig:overview} (e) and (h). During the initial phase, the jet developed into a long cylindrical shape with many helical bright plasmas on its surface (see \nfig{fig:overview} (c)-(f)). Subsequently, as the whole helical body was rising higher, the jet displayed a highlight untwist magnetic field lines (see \nfig{fig:overview} (f)). For more details of the jet's evolution process, please refer to the animation of \nfig{fig:overview}.

Upon comparing observation of the jet from the AIA 304 \AA\ channel with it from 131 \AA\ channel observations, one can note that AIA 131 \AA\ channel images had a more simple bright structure along the line-of-sight. It facilitates us to use them to track the jet's rotation process. To reveal the kinematics of the jet more clearly, we established a coordinate system $x'-y'$ with its $y'-$axis along the jet's center axis to \xp{track} the rotation motion, as shown in \nfig{fig:overview} (g). Its origin coordinate $o'$, $[a,b]$, is $[937,228]$, and the rotation angle, $\theta$, is $-(0.93+\pi/2)$. \nfig{fig:tdp} (a1) shows the relative position of the coordinate system $x'-y'$, with a rotation angle $\theta$, to the solar arcsec $x-y$ coordinate system. We track a specific moving bright feature (MBF, \zxp{whose location has been marked by a white arrow in panel AIA 131 \AA\ of the animation of \nfig{fig:overview}}) during 08:42:08\,UT$-$08:43:20\,UT in the solar arcsec coordinate system $x-y$.  Their locations relative to the $x'-y'$ coordinate system, transformed using Equations \ref{eq1} and \ref{eq2}, are overlaid on it marked with ``+'' in the inset of \nfig{fig:tdp} (a1).

\begin{equation}\label{eq1}
	\left\{
	\begin{aligned}
		x=a+x'\cos\theta-y'\sin\theta	\\
		y=b+x'\sin\theta+y'\cos\theta
	\end{aligned}
	\right.
\end{equation}

\begin{equation}\label{eq2}
	\left\{
	\begin{aligned}
		x'=y\sin\theta-b\sin\theta-a\cos\theta+x\cos\theta \\
		y'=a\sin\theta-x\sin\theta+y\cos\theta-b\cos\theta	
	\end{aligned}
	\right.
\end{equation}
Based on the coordinates in the $x'-y'$ coordinate system, we present the temporal evolution of the $x'-$ and $y'-$components in Figure \ref{fig:tdp} (a2) and (a3), respectively. By analyzing the distributions of these two components, we can easily find that the motion of the MBF exhibits a spiral pattern around \xp{the $y'-$axis, i.e., the jet's} center axis. To characterize this motion, we fit a \zxp{cosine function, \zxp{$x'=rsin(\frac{2\pi}{T}(t-\sigma))$}, to the transverse motion in the $x'$ direction, assuming that the radius of the jet dose not change during selected period,} and a linear function to the radial motion in the $y'$ direction. Our fitting yields the following parameters for the jet: a period of $T =120\pm8$\,s, an angular speed of $\omega=0.052$ rad\,s$^{-1}$ (or $3^\circ$\,s$^{-1}$), a rotation radius of $r=6.5\pm0.4$\,Mm, a linear velocity of $v_l=340\pm$\speed{12}, and a radial speed of $v_r=375\pm$\speed{15}. Also, we select five cuts, marked with S1-S5 in \nfig{fig:overview} (c), to study the kinematic of the jet in the transverse and radial directions, and the relevant time-distance stack plots are displayed in \nfig{fig:tdp}. Consequently, the speed of the helical structure is in the range of 150-\speed{160} through fitting the bright stripes on the time-distance stack plots using a linear function (see \nfig{fig:tdp} (b1)-(c3)), while its radial speed is 405$\pm$\speed{23} (see \nfig{fig:tdp} (d)). Compared with \xp{these} two methods mentioned above, we can find that the obtained radial speeds are similar, \speed{375} and \speed{405}, while the transverse speeds have a relatively significant difference, \speed{340} and \speed{150}. The disparity in transverse speeds is reasonable given that  \xp{it} derived from the time-distance stack plots represent \xp{projection} speeds on the sky plane and do not consider the movement of the MBF along the surface of the cylindrical jet. Additionally, the diameter of the jet has shown a substantial increase from approximately 14\,Mm to 22\,Mm (refer to \nfig{fig:tdp} (b2)). It suggests that the jet experienced significant expansion during its eruption. It is important to note that this expansion phenomenon is exclusive to the side far to the solar surface. Conversely, on the side closer to the solar surface, the \xpz{column plasma density} of the jet has exhibited a considerable rise compared to \xp{that} observed on the sides away from the solar surface, \zxp{which can be verified by the obvious difference brightness between them in \nfig{fig:tdp} (b1)-(c3)}. This \xpz{column plasma density} disparity on both side of the jet can be attributed to the mutual compression between the jet and the solar surface.

\subsection{Kinematics of the QFP wave trains}

\begin{figure}
	\centering
	\includegraphics[width=.75\linewidth]{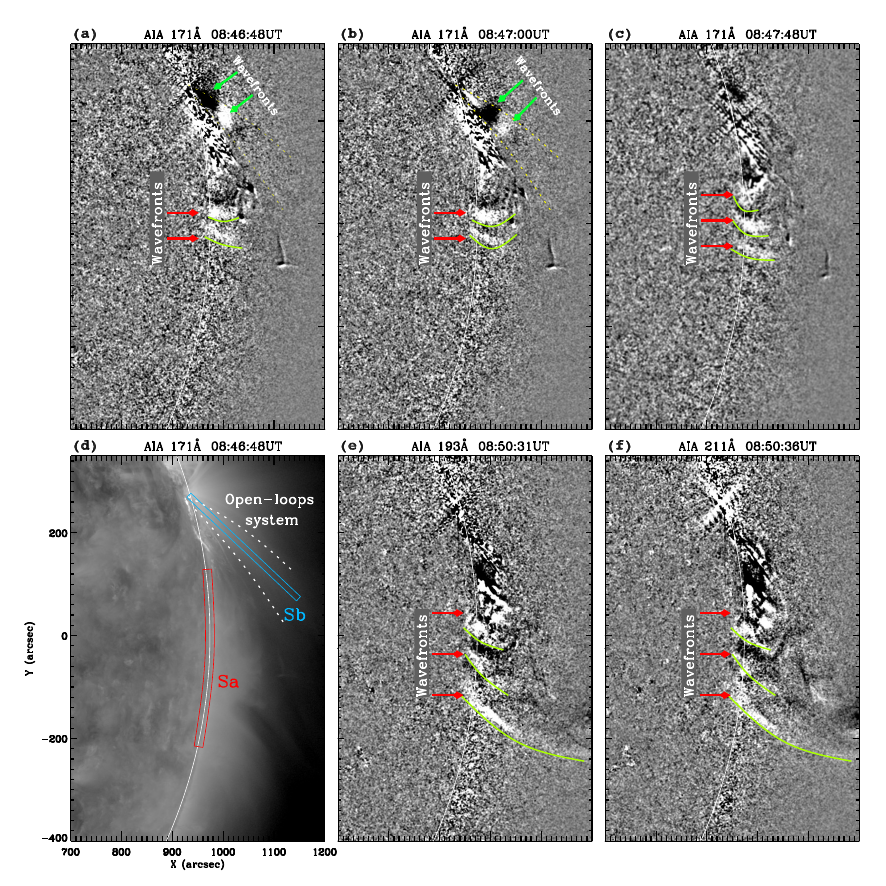}
	\caption{The running-difference images in SDO/AIA 171 \AA, 193 \AA\ and 211 \AA\ show the snapshots of the wave train at different times, in which the green and red arrows indicate the wavefronts of the narrow and broad QFP wave trains, respectively. Meanwhile, the wavefronts of the broad QFP are also highlighted by the green curve in panels (a)-(c) and (e)-(f). The rectangles marked Sa and Sb overlaid on the SDO/AIA 171 \AA\ \xp{raw image} are used to reconstruct the time-distance plots in \nfig{fig:wavelet}, while the white dotted line sketch the boundary of the open-loops system. \zxp{An animation of this image is available. The animation covers 08:35\,UT-9:05\,UT with a 12\,s cadence. The animation is 5\,s.}
	\label{fig:evolution}}
\end{figure}

Here, we mainly analyze the time sequence of the AIA 171 \AA, 193 \AA, and 211 \AA\ running-difference images to the overall evolutionary process of the wave trains. \xpz{In the technique of running difference, each image in a sequence of images is subtracted from the previous 12\,s image to enhance moving features.} As shown in \nfig{fig:evolution}, during the violent eruption launched two different wave trains: multiple successive wavefronts with an \zxp{angle}  about 15$^\circ$, observed from the perspective of the SDO, propagating along the solar limb, others, with an angle about \xp{5}$^\circ$ running along the funnel-like open-loops system.\footnote{{According to the statistics by \citep{2022SoPh..297...20S}, the typically angular width of the narrow and broad QFP wavefront are in the range of 8$^\circ$-80$^\circ$ and 90$^\circ$-360$^\circ$, respectively. The angular width, $15^\circ$, of broad QFP wavefronts present here significantly out its typical range. That is due to the affect of projection effect, resulting only observing part of the wavefront from the view of SDO.} } These two wave trains are marked with red and green arrows in \nfig{fig:evolution}, respectively. The images demonstrate the presence of the broad wave train, which, much like the classical EUV wave, is prominently visible in 171 \AA, 193 \AA, and 211 \AA\ channels, with a particularly distinct appearance in the 193 \AA\ and 211 \AA\ channels. On the other hand, the narrow wave train is solely observable in the AIA 171 \AA\ channel. These characteristics are consistent with previous reports about them \citep{,2011ApJ...736L..13L,2012ApJ...753...53S,2022ApJ...941...59Z,2022A&A...659A.164Z,2022ApJ...930L...5Z}. \xpz{Particularly, the first wavefront of the broad wave train appears much longer than the following ones because the unwinding motion is a spatially widening spiral, resulting in the front part of the jet compressing the plasma more pronounced on the side near the solar surface. In this process, the closer to the solar surface side, the more pronounced the jet squeezing of the solar atmosphere, which results in the intensity of the wavefront becoming progressively stronger as it approaches the solar surface. This can be verified from \nfig{fig:evolution} (e) and (f). Additionally, the fact that the second and third wavefronts are weaker than the first one leads to their length of manifestation appearing to be much shorter than that of the leading wavefront.} It is noteworthy that the narrow QFP wave \zxp{became pronounced since 08:45:08 UT near the flare kernel, However, the amplitude is smaller than that after 08:46:44\,UT. The broad QFP wave was observed at a considerable distance of approximately 100\,Mm, initiated at 08:44:56\,UT (see the animation of \nfig{fig:evolution}).} \xpz{Notably, we obtained the same QFP structures after increasing the time resolution for reconstructing running-difference images from 12\,s up to 24\,s and 36\,s, indicating that the wave signal was not affected by the stroboscopic effect. }

To analyze the kinematics of the broad  \xpz{ and narrow}  wave trains, we reconstructed the time-distance stack plots using the AIA 171 \AA\ and 193 \AA\ running-difference images along the paths Sa and Sb shown in \nfig{fig:evolution} (d). The time-distance stack plots are displayed in \nfig{fig:wavelet} (a1)-(a3), in which one can find that there exist multiple wavefronts with a speed of 370$\pm$\speed{16} and an acceleration -0.43$\pm$\acc{0.01}. It is worth noting that the wavefronts gradually become diffuse after propagating about 70 Mm in the AIA 171 \AA\ channel (as shown in \nfig{fig:wavelet} (a1)), resulting in wavefronts features that are increasingly difficult to be distinguished. This phenomenon is similar to the classical EUV wave that is best seen in 193 \AA\ and 211 \AA\ channels \citep{2014SoPh..289.3233L}.  In contrast to the broad QFP wave train propagating along the solar limb, the \xp{observed} narrow QFP wave train confined in the loops system and only to be observed in 171 \AA\ channel, propagated with a higher speed of over \speed{1000} (see \nfig{fig:wavelet} (a3)), which is consistent with that reported by previous researchers \citep{2018MNRAS.477L...6S,2021ApJ...908L..37M,2022A&A...659A.164Z}. The speed and acceleration of the broad QFP wave train propagating along the solar limb are in the range of the classical EUV wave \citep{2013ApJ...776...58N}, but significantly lower than that of the narrow wave train (dominate speed is over \speed{1000} as reported by \cite{2022SoPh..297...20S}). These features indicate that the broad wave train present here, similar to that reported by \cite{2022A&A...659A.164Z,2022ApJ...930L...5Z}, is a rare broad \xp{QFP} wave train.

\begin{figure}
	\centering
	\includegraphics[width=.75\linewidth]{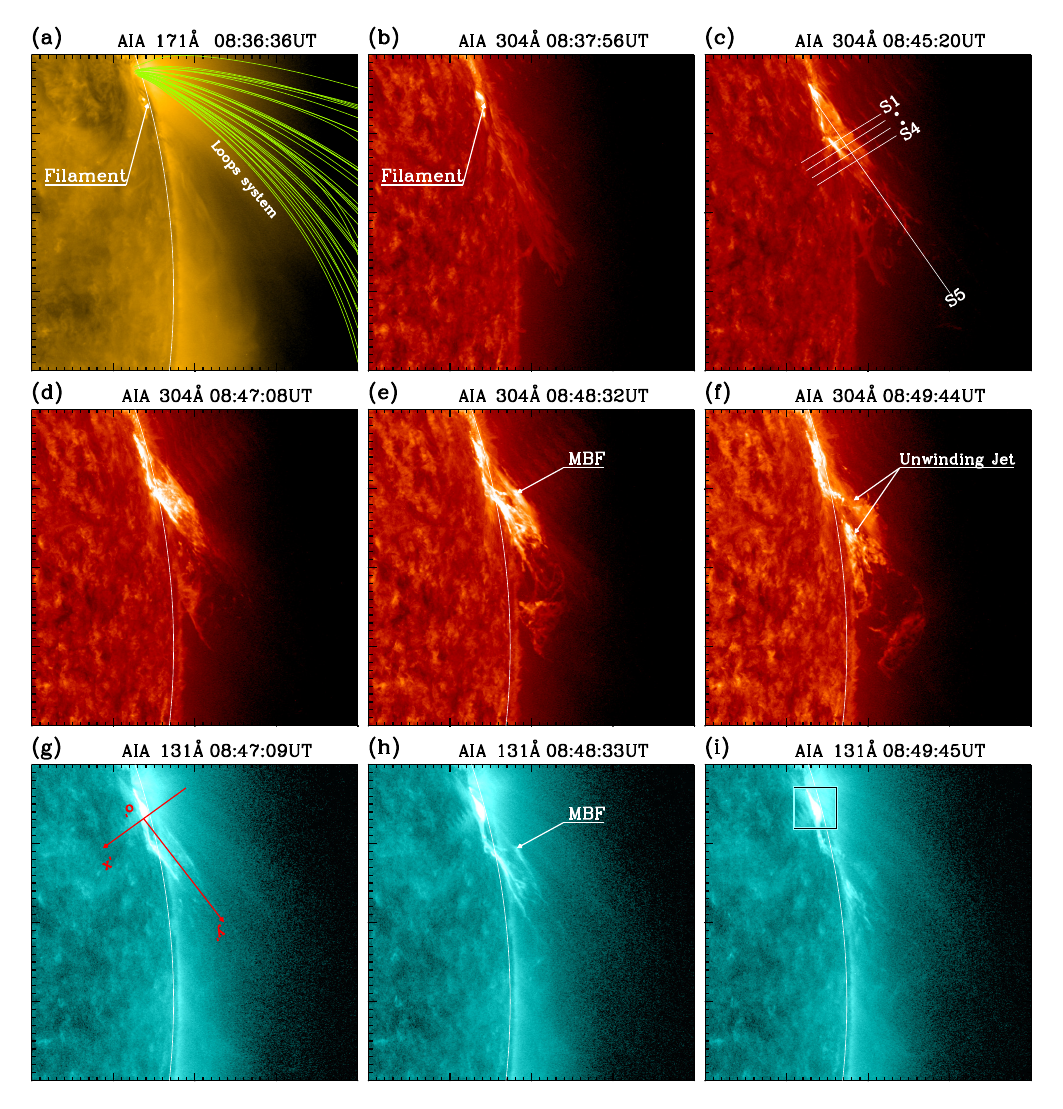}
	\caption{Panel (a) shows the initial coronal condition of the eruption source region using the SDO/AIA 171 \AA\ \xp{raw image}, where the green curve lines overlying the filament are the extrapolated open magnetic-field lines obtained using the PFSS model. Panels (b)-(i) display the evolution of the jet with the SDO/AIA 304 \AA\ and 131 \AA\  raw image. The white lines marked S1-S5 in panel (c) represent the positions to obtain time-distance stack plots to trace the evolution of the jet in perpendicular and parallel directions to the jet's central axis. Two perpendicular coordinate axes labeled with $x$\uu~and $y$\uu~overlaid in panel (g) are used as a coordinate system to trance the moving bright features, which are also plotted in \nfig{fig:tdp} (a1). An animation of this image is available. The animation covers 08:35\,UT-08:55\,UT with a 12\,s cadence. The animation is 3\,s. 
	\label{fig:overview}}
\end{figure}

\begin{figure}
	\centering

	\includegraphics[width=.75\linewidth]{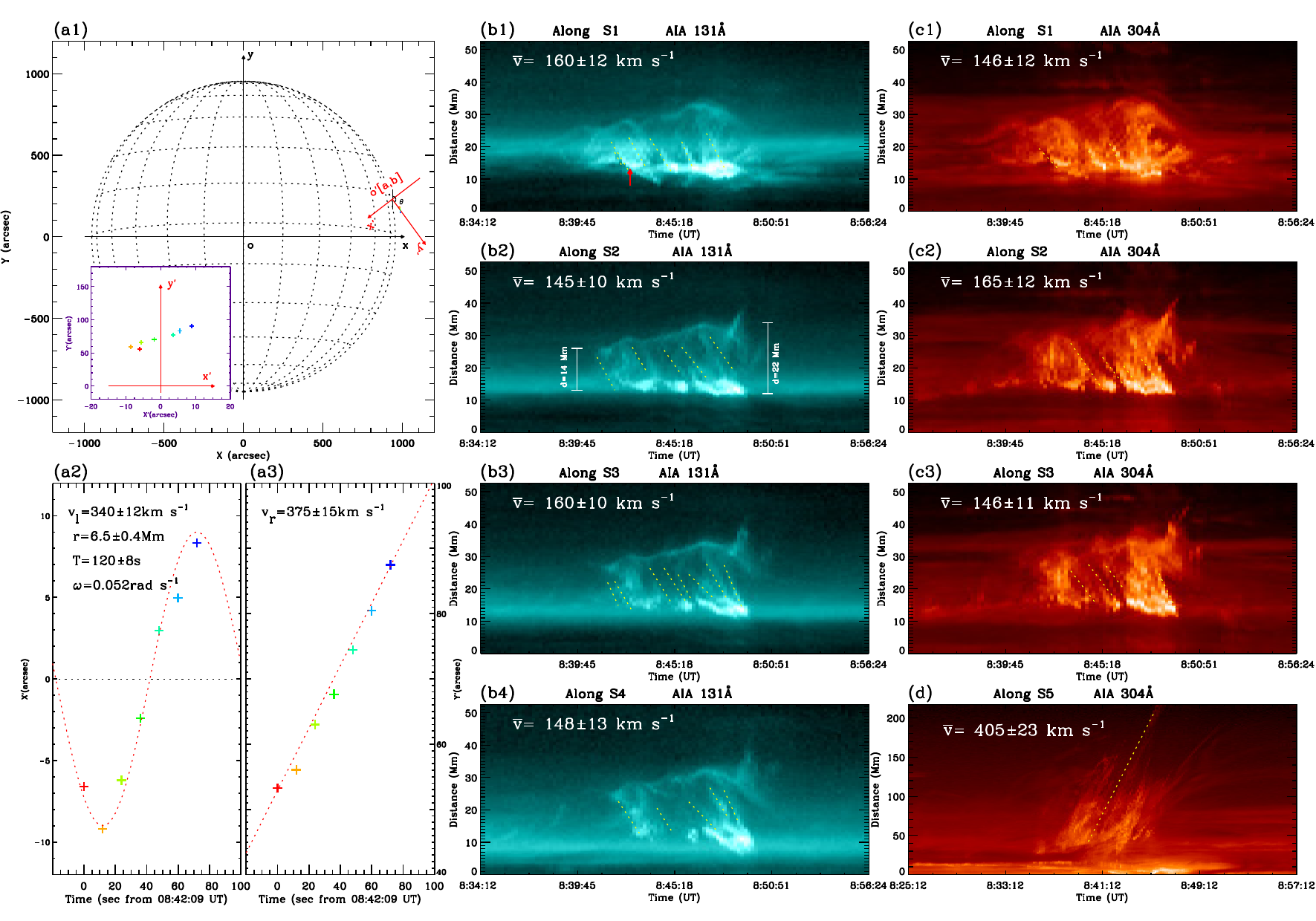}
	\caption{ Panel (a1) shows the coordinate system $x'-y'$ relative to the arcsec coordinate system $x-y$. The color ``+'' symbols in the inset of panel (a1) represent the variations of the location coordinates of the moving bright features relative to the $x'-y'$ coordinate system. Panels (a2) and (a3) show the evolution of moving bright features in the $x'$- and $y'$-directions, respectively. The fitted parameters are list on the corresponding panels. Panels (b1) -(d) are the time-distance stack plots reconstructed along the slices S1-S5 in \nfig{fig:overview}, using the AIA 131 \AA\ (b1-b4) and 304 \AA\ (c1-d) \xp{raw images}. The ridges' slope represents the moving material's speeds, and their speeds obtained by linear fitting are listed in each panel. \xp{Errors in parameters such as speed, period, etc are estimated by making the fit ten times.} 
		\label{fig:tdp}}
\end{figure}

\begin{figure}
	\centering
	\includegraphics[width=.75\linewidth]{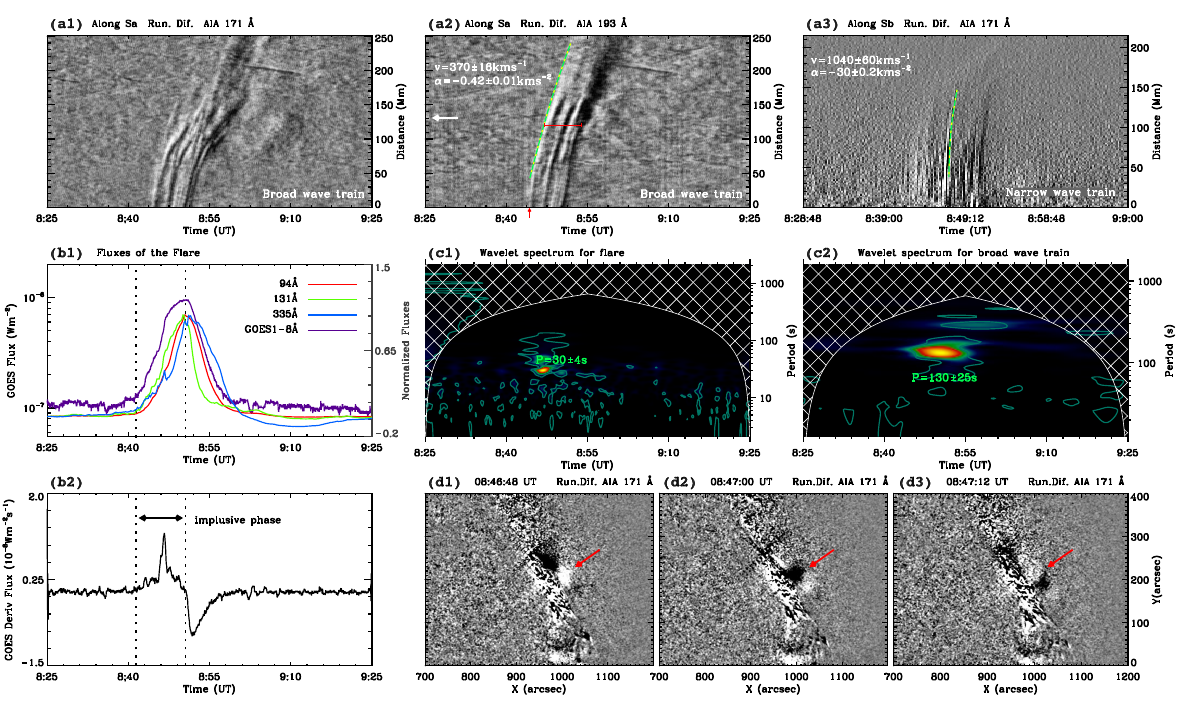}
	\caption{\xpz{Panels (a1)-(a3) are the time-distance stack plots obtained from the AIA 171 \AA\ and 193 \AA\ running-difference images along the slice Sa and Sb shown in \nfig{fig:evolution} (d). The white arrows in panels (a2) point to the positions where the intensity profile is used to analyze the periodicity of the wave trains. Panel (b1) display the GOES 1-8 \AA\ X-ray flux and normalized AIA 94 \AA, 131 \AA\ and 335 \AA\ light flux curves within the eruption source region marked with the black box in \nfig{fig:overview} (i). Panel (b2) is the derivative curve of the GOES 1-8 \AA\ flux. Panel (c1) and (c2) respectively show the wavelet spectrum of the flare and broad QFP wave train, and their corresponding periods are listed in the figures. \xp{The main periods and their errors are determined by the peak and full width at half maximum (FWHM) of the corresponding global wavelet power spectra. The dark blue curves in wavelet spectrum outline the 95\% significance level, while the cross-hatched areas in wavelet spectrum indicate the cone of influence region due to the edge effect of the data. Panels (d1)-(d3) are three snapshots to show the evolution of the narrow QFP wave train, where the red arrow in each panel is located at a specific position.}}
		\label{fig:wavelet}} 
\end{figure}

The quasiperiodic pulsations (QPPs) can be responsive to the quasiperiodic energy release process \citep{2007LNP...725..221N,2009SSRv..149..119N,2020A&A...639L...5L,2022ApJ...931L..28L,2024RvMPP...8....7Z}, which have the potential to trigger either a narrow wave train \citep{2011ApJ...736L..13L,2012ApJ...753...53S,2022ApJ...941...59Z} or broad wave train \citep{2021SoPh..296..169Z,2022ApJ...930L...5Z,2022A&A...659A.164Z}, via magnetic reconnection. Comparing the AIA 94 \AA, 131\AA\ and 335 \AA\ light curves in \nfig{fig:wavelet} (b1) measured from the flaring kernel region (see the black box in \nfig{fig:overview} (i)), one can find that they have a similar trend with the GOES 1-8 \AA\ flux (with an impulsive phase about 8:41\,UT to 8:51\,UT, see \nfig{fig:wavelet} (b2)), suggesting that no violent eruption in other regions during the present event.  \xp{According to the Neupert effect \citep{,1968ApJ...153L..59N,2024SoPh..299...57L}, the time-derivative of soft X-ray fluxes can be viewed as a proxy of the corresponding hard X-ray fluxes that reflects the variation of non-thermal particles accelerated by the magnetic reconnection in flares. The detrend time-derivative of GOES 1-8 \AA\ flux obtained from the intensity profile by subtracting its smoothed intensity profile with 60\,s boxcar, which enhances periods shorter than 60\,s, thus highlighting short-period periodicity features in the wavelet spectrum \citep{2013SoPh..284..559K,2021ApJ...921..179L}.}  \xpz{Additionally, the time cadence of the GOES data is 2\,s. \zx{Therefore, the GOES data is suitable for detecting the periods of the flare QPPs longer than 6 seconds with the wavelet analysis} \citep{1998BAMS...79...61T}  that has been widely utilized to analyze the periodicity of time-dependent one-dimensional data \citep{2015A&A...581A..78Z,2016ApJ...832...65Z,2017MNRAS.471L...6L,2018ApJ...868L..33L,2020ApJ...893....7L,2020ApJ...893L..17L,2022ApJ...937L..21Z,2024ApJ...970...77L}.} Using the detrend time-derivative of GOES 1-8 \AA\ flux as input and selecting the Morlet function as the mother function, the wavelet spectrum shows that the flare period is about 30$\pm$4\,s, as shown in \nfig{fig:wavelet} (c1). To reveal the period of the broad QFP wave train, we extracted the intensity profile at the positions marked with white arrows in \nfig{fig:wavelet} (a2), and the wavelet spectrum is shown in \nfig{fig:wavelet} (c2). Clearly, the main periods of the broad QFP wave train was about 130$\pm$25\,s.  \xpz{For the broad QFP wave train, the smoothing time window width used here is 4$\times$60\,s. We get a similar result when using 0.4$\times$60\,s, 1$\times$60\,s, 2$\times$60\,s, 3$\times$60\,s as the smoothing time window width, indicating the period of the broad QFP wave not an artifact of the smoothing procedure.}
\xpz{We also check the periods of the broad QFP wave by another method: The red line in \nfig{fig:wavelet} (a2) marks the duration of the three wavefronts, which is about 7 minutes. Therefore, we can get its period to be about $7\times 60 \div3=140$\,s. This result is similar to that obtained from the wavelet analysis, indicating that the wavelet analysis should be reliable.}  \zx{Combining the animation of \nfig{fig:evolution}} and the time-distance stack plot of the narrow QFP wave train displayed in \nfig{fig:wavelet} (a3), we can find that the frequency of the narrow QFP wave train was significantly high than that of the broad QFP wave. In \nfig{fig:wavelet} (d1)-(d3), we can see that a wavefront appeared at the red arrow position at 08:46:58 UT, 12\,s later, at 08:47:00 UT the dimming region followed the wavefront reached the red arrow position. Then, at 08:47:12 UT, the second wavefront appeared at the red arrow position. Therefore, we can infer that the period of the narrow QFP wave train is about 24\,s, which is in good agreement with that of the flare QPPs. \zx{However, we are not able to prove this period using the wavelet analysis because of the low time resolution (12 s) of the SDO data}. The period of the broad wave train is much larger than that of the flare' QPPs but shares a common period, about 2\,minutes, with that of the \xp{unwinding} jet. The close period relationship between the unwinding jet and the broad QFP wave train suggests that the broad \xp{QFP} wave train should have a tight connection with the successive expansion and untwist of the rotation jet.  Additionally, according to the statistics by \cite{2022SoPh..297...20S}, the broad QFP wave train is always associated with the flare class exceeding the C-class. This correlation arises from the higher energy threshold required to trigger a broad wave train compared to a narrow wave train, leading to the fact that the B-class flare reported here can not trigger the broad wave train instead of the narrow wave train. This result is supported by the fact that the energy flux density of the broad wave train is typically higher than that of the narrow wave train, as indicated in the statistics by \cite{2022SoPh..297...20S} (where the energy flux density of the broad and narrow waves ranges from 10-19 $\times10^5$\,erg\,cm$^{-2}$\,s$^{-1}$ and 0.1-4 $\times10^5$\,erg\,cm$^{-2}$\,s$^{-1}$, respectively). \xp{For the narrow QFP wave train,  we believe that, as the previous results \citep{2011ApJ...736L..13L,2012ApJ...753...53S}, the flare's QPPs is their driving source due to the close temporal and period relationship between this two.}

\section{Discussion and conclusions} 
\label{se:discussion}

	In this work, we studied the simultaneous appearance of two scarce \xp{QFP wave trains, each characterized by multiple wavefronts}. The event associated with a unwinding jet eruption on 2011 January 27. The broad QFP wave train was observed to propagate along the solar disk limb at a speed of 370$\pm$\speed{16}. In contrast, the narrow QFP wave train, confined within a funnel-shaped loops-system that overlaid the jet, propagated with a speed of 1040$\pm$\speed{60}. \xp{Upon analyzing the measured parameters, our result suggests that the broad QFP wave train reported here is not predominantly associated with flare QPPs}; instead, it appears to be triggered by an unwinding jet, contrary to earlier studies. On the contrary, \zxp{the released energy behind the flare QPPs} excited the narrow QFP wave train.

To date, the number of observed broad QFP wave events is quite limited, with up to 8 such events identified. A comprehensive explanation for their driving mechanism, however, remains elusive. More than twenty years ago, \cite{2002ApJ...572L..99C} proposed that the broad QFP wave train could be caused by the successive stretching of the magnetic field lines during the eruption of the filament. However, recent research by \cite{2021ApJ...909...45Y} using a 2.5D MHD simulation suggests that turbulence plays a crucial role in generating the broad QFP wave train around the CME. Furthermore, \cite{2024ApJ...962...42H} offers a new perspective on understanding the triggering of broad QFP wave trains. They propose that the leakage of internal disturbances in the magnetic rope acts as a waveguide, forming a multiple wavefront structure as a reasonable mechanism for wave train generation. The case in the current study cannot or is difficult to provide observational evidence for supporting these numerical simulations. \cite{2012ApJ...753...52L}, first reported the broad QFP wave train, believe that the broad QFP wave train should be driven by the periodic energy release of the associate flare because the wave train shares a common periodic with the flare. This view is supported by the subsequent observational event and numerical simulations; for example, \cite{2022A&A...659A.164Z,2022ApJ...930L...5Z} find these wave trains share a common period with the accompanying flare. In the numerical experiment conducted by \cite{2021ApJ...911L...8W}, a series of wavefronts with a dome-shape excited by the energy release associated with the flare. However, in the event reported here, the period of the accompanying flare does not align with that of the broad wave train but is consistent with that of the narrow wave train. Thus, we propose that the flare only excited the narrow QFP wave train. Previous studies have suggested that jets can drive coronal waves in front of them \citep{2018ApJ...861..105S,2022ApJ...926L..39D}, equivalent to the piston-driven scenario in a shock tube: once the shock wave detaches with the piston (jet), it can continue propagation freely \citep{2013SoPh..286..509L}. Particularly, \cite{2018ApJ...861..105S} identified four homologous EUV waves originating from the same active region. The findings indicate that those recurrent waves were driven by the recurrent jets. In contrast to the that scenario, the jet in the present case did not erupt intermittently; instead, it exhibited a pronounced periodic rotational motion during its eruption. By tracking a bright feature moving helically in the jet, we find that its unwinding period is about 120\,s. This period is consistent with that of the broad QFP wave train. \xp{On the other hand}, the onset time of the narrow QFP wave train and flare QPPs was essentially the same, occurring at 8:46\,UT (see the animation of \nfig{fig:evolution}). However, the initiation time of the broad QFP wave train was approximately 08:44\,UT (refer to the red arrow in \nfig{fig:wavelet}(a2)), coinciding with the interaction between the unwinding structures of the jet and the solar surface (refer to the red arrow in \nfig{fig:tdp}(b1)) but preceding the narrow QFP wave train and flare by approximately 2\,minutes. Based on this temporal and period relationship, we propose that the broad QFP wave train \xp{should not be triggered by the flare QPPs, but is the jet's rotation motion. The jet' unwinding is }also one type of periodic energy release of the magnetic twist stored in the filament into the outer corona \citep{1986SoPh..103..299S,1996ApJ...464.1016C,2013RAA....13..253H}. The reported broad QFP here is potentially formed as a result of the untwisting of the filament via jet. Jets involving filament eruptions are generally called blowout jets \citep{2010ApJ...720..757M,2015Natur.523..437S} or breakout jets \citep{2017Natur.544..452W} in historical literature. No matter in which jet model, as the filament in the jet base continuously ascends, interchange reconnection between the ambient field and filament is expected to occur. Once such interchange reconnection occurs, the twist of the filament ultimately would relax into the ambient field and form the unwinding jet \citep{2011ApJ...735L..43S,1996ApJ...464.1016C,2009ApJ...691...61P,2023ApJ...942...86Y,2024ApJ...964....7Y}.

We find that many jets with highly rotational motion are not accompanied by broad QFP wave train, such as the jets reported by \cite{2011ApJ...735L..43S}, \cite{2013RAA....13..253H}, \cite{2015SoPh..290.2857L} and \cite{2021ApJ...911...33C}. We speculate that it may be due to the incline angle of the jet's axis relative to the solar surface being too large; for example, the jets reported by \cite{2011ApJ...735L..43S} and \cite{2013RAA....13..253H}, they were almost near to $90^\circ$. We can get inspiration from Zheng's research result for the explanation of the origin of the Moreton wave in the Chromosphere \citep{2023ApJ...949L...8Z}: only when the direction of the eruption is \xp{incline} enough relative to the solar surface can it cause effectively disturb the solar corona. This can be verified \xp{by the appearance position of the wave signal which is only in the southward direction consistent with the jet inclination in the current event}. We hypothesize that, apart from the \xp{incline} angle, the expansion occurring during jet untwisting could also be a crucial factor in initiating the broad QFP wave train in the current event because it can facilitate the efficient release of stored magnetic twist energy.

Studying the coronal waves is of particular importance in solar physics since they play an essential role in coronal heating \citep{2019ApJ...882...90L,2022ApJ...941...59Z}, diagnosing the magnetic field strength \citep{1970PASJ...22..341U,1984ApJ...279..857R} and detected the information inside the coronal hole using the lensing effect \citep{2024NatCo..15.3281Z}. The single impulsive lateral expansion of the CME, an explanation for the driven mechanism of the classical large-scale coronal wave, is not adapted to explain the origin of these broad waves characterized by multiple wavefronts. Consequently, \xp{We anticipate that a unified explanation of the QFP wave train will materialize in the future. Of course,  it is also possible that triggering mechanisms itself is diverse. As the Sun enters the solar cycle 25, the sun \zxp{is reaching its maximum} \citep{2023SCPMA..6629631C}. This is conducive to combining the observation by SDO and new solar observatories recently launched by China, such as ASO-S \citep{2019RAA....19..156G,2023SoPh..298...68G}, CHASE \citep{2022SCPMA..6589602L} and SUTRI \citep{,2023RAA....23f5014B}, to carry out more comprehensive study on the trigger mechanism of the QFP wave trains, especially, the broad QFP wave train. }
 
\section{Acknowledgments}
We want to thank the anonymous referees for the many valuable suggestions and comments for improving the quality of this paper, and we are thankful to the teams of SDO and GOES for providing excellent data. This work is supported by the Natural Science Foundation of China (12303062), the Sichuan Science and Technology Program (2023NSFSC1351), and the Project Supported by the Specialized Research Fund for State Key Laboratories. Y.D.S is supported by the Natural Science Foundation of China (12173083) and the Yunnan Science Foundation for Distinguished Young Scholars (202101AV070004). Coauthor A.A.I would like to thank the Researchers Supporting Project (Grant No. RSPD2024R993), King Saud University, Riyadh, Saudi Arabia. We gratefully acknowledge ISSI-BJ for supporting the international team ``Magnetohydrodynamic wave trains as a tool for probing the solar corona''.  We also acknowledge Sichuan Normal University Astrophysical Laboratory Supercomputer for providing providing the computational resources. \xp{The Wavelet\footnote{\url{https://paos.colorado.edu/research/wavelets/}} software was provided by C. Torrence and G. Compo.}

\vspace{5mm}
\end{CJK*}
\end{document}